\documentclass[12pt]{iopart}
\usepackage{graphicx}
\usepackage[usenames,dvipsnames]{color}
\usepackage{amsfonts}
\usepackage{hyperref}
\usepackage{epigraph}
\usepackage{booktabs}
\usepackage{multirow}
\usepackage{placeins}

\usepackage[backend=biber,style=ieee,eprint=false]{biblatex}
\DeclareUnicodeCharacter{2009}{\,}

\newcommand{\mmps}{\,\mathrm{m}^2/\mathrm{s}}
\newcommand{\T}{\,\mathrm{T}}
\newcommand{\m}{\,\mathrm{m}}

\newcommand{\MW}{\,\mathrm{MW}}

\newcommand{\atoms}{\,\mathrm{atoms}/\mathrm{s}}

\newcommand{\eV}{\,\mathrm{eV}}

\newcommand{\Pin}{P_{\mathrm{in}}}
\newcommand{\Dpuff}{D_{\mathrm{puff}}}
\newcommand{\Npuff}{N_{\mathrm{puff}}}
\newcommand{\Dcore}{D_{\mathrm{core}}}

\newcommand{\teot}{T^\mathrm{ot}_{e,\mathrm{sep}}}
\newcommand{\teomp}{T^\mathrm{omp}_{e,\mathrm{sep}}}
\newcommand{\ot}{\mathrm{ot}}
\newcommand{\omp}{\mathrm{omp}}

\DeclareUnicodeCharacter{2009}{\,}

\begin{document}

\title[SOLPS-NN]{Deep-Learning based surrogate models for plasma exhaust simulations - SOLPS-NN}

\author{S.~Dasbach$^{1}$, S.~Brezinsek$^{2,3}$, Y.~Liang$^{2}$, D.~Reiser$^{2}$, S.~Wiesen$^{1,4}$}

\address{$^1$DIFFER - Dutch Institute for Fundamental Energy Research, De Zaale 20, 5612 AJ Eindhoven, Netherlands}
\address{$^2$Forschungszentrum Jülich GmbH, Institute of Fusion Energy \& Nuclear Waste Management - Plasma Physics, 52425, Jülich, Germany}
\address{$^3$Heinrich Heine University Düsseldorf, Faculty of Mathematics and Natural Sciences, 40225, Düsseldorf, Germany}
\address{$^4$EUROfusion, Boltzmannstr. 2, Garching 85748, Bavaria, Germany}
\ead{s.f.w.dasbach@differ.nl}
\vspace{10pt}
\begin{indented}
\item[]February 2026
\end{indented}

\begin{abstract}
Accurate models of the scrape-off layer are required for the design and operation of tokamak fusion reactors. Scrape-off layer simulations are computationally expensive, difficult to operate and suffer from numerical instabilities. A potential remedy comes in using machine learning models trained on simulations for fast and easy to use predictions. We present a such candidate surrogate model - named SOLPS-NN - to provide recommendations for the methods to construct it. Based on a large dataset of several thousand SOLPS-ITER simulations with reduced neutral fidelity, a variation of machine learning models with differing architectures and scopes are tested. The evaluation shows that simple fully connected neural networks are a suitable architecture. It is demonstrated that the whole spatial domain can be predicted at once, but that it is easier to achieve high accuracy by employing independent models for different observables. The presented surrogate model with reduced neutral fidelity is sufficient to predict access to detachment with trends similar to experiments. A small dataset of higher fidelity ITER baseline SOLPS-ITER simulations is used to (re-)train surrogate models. The smaller extent of the ITER dataset allows for achieving much more accurate predictions. Transfer learning from the previous surrogate model works but has no direct benefits over training a new model from scratch. Future efforts should focus on discovering the potential and the methods for models utilizing simulations with mixtures of fidelity.
\end{abstract}

%
\vspace{2pc}
%
%
%
%
\newpage
\section{Introduction}\label{sec:intro}
The Tokamak Scrape-Off Layer (SOL) is critical for the overall performance of tokamak devices \cite{wesson_tokamaks_2004,zohm_physics_2013}.
It serves as the primary channel for heat and particle transport out of the plasma.
This affects the overall energy balance and drives the complex interactions between the hot plasma and the material surfaces inside the tokamak.
Understanding and managing the processes within the SOL is imperative for safe divertor operation while maintaining the required boundary conditions for the stability and performance of the confined plasma \cite{wenninger_demo_2014,pitts_physics_2019,zohm_eu_2021}.
Simulation codes such as SOLPS-ITER \cite{wiesen_new_2015} contain sufficiently sophisticated physics models but suffer from prolonged convergence times \cite{kukushkin_finalizing_2011,wiesen_plasma_2017}.
Additionally the SOL is sensitive to numerous parameters, some of which remain a priori unknown or subject to significant uncertainties \cite{iter_organization_iter_2018}. 
This prevents rapid design studies, automatic optimization or coupling with simulations for the pedestal or plasma-wall interaction (PWI).
To deal with similar problems, many fields in science and engineering but also in nuclear fusion employ surrogate models \cite{sun_review_2019,pruett_creation_2016,zhang_dnn-assisted_2020,van_de_plassche_fast_2020,preuss_gaussian_2017,gopakumar_image_2020,zhu_data-driven_2022,gopakumar_plasma_2023,holt_tokamak_2024,zhu_latent_2025,luo_neural_2025a,luo_neural_2025b}.
Surrogate modeling uses machine learning techniques to approximate computationally expensive simulations, by creating a simpler faster-to-evaluate model instead. This model can than be used in optimization, design exploration and uncertainty quantification.
Informed by simulations surrogate models can be applicable in a parameter space extending beyond existing experiments.
The central challenge in surrogate modeling is managing the trade-off between model accuracy and computational efficiency.
This involves selecting an appropriate model type, efficient simulation data generation and ensuring the accuracy in capturing key underlying phenomena.
\\
The work presented here constitutes a summary of the dissertation \cite{dasbach_surrogate_2025}.
As such the work is a continuation and extension of the work already published in \cite{dasbach_towards_2023,wiesen_data-driven_2024}. The goal of this study is to develop and compare models that extend the capabilities of previous models by providing results for the whole SOL domain and multiple observables, to showcase some of the results obtainable with these models and demonstrate strategies to mitigate the discrepancy between these models and higher fidelity simulations. This publication also acts as a reference paper for the developed SOLPS-NN surrogate model.
Section \ref{sec:dataset} gives a brief introduction into the simulation dataset, while Section \ref{sec:2D} compares models predicting the electron temperature in the SOL. Sections \ref{sec:multi} and \ref{sec:loop} compare different models and strategies to incorporate multiple observables. Section \ref{sec:physics} uses the final model to predict the impurity concentration necessary for detachment across several machines. Finally Section \ref{sec:transfer} compares the model predictions to higher fidelity simulations and tests calibration strategies.

\newpage
\FloatBarrier
\section{The fluid neutral dataset}\label{sec:dataset}
\begin{figure}
    \centering
    \includegraphics[width=\textwidth]{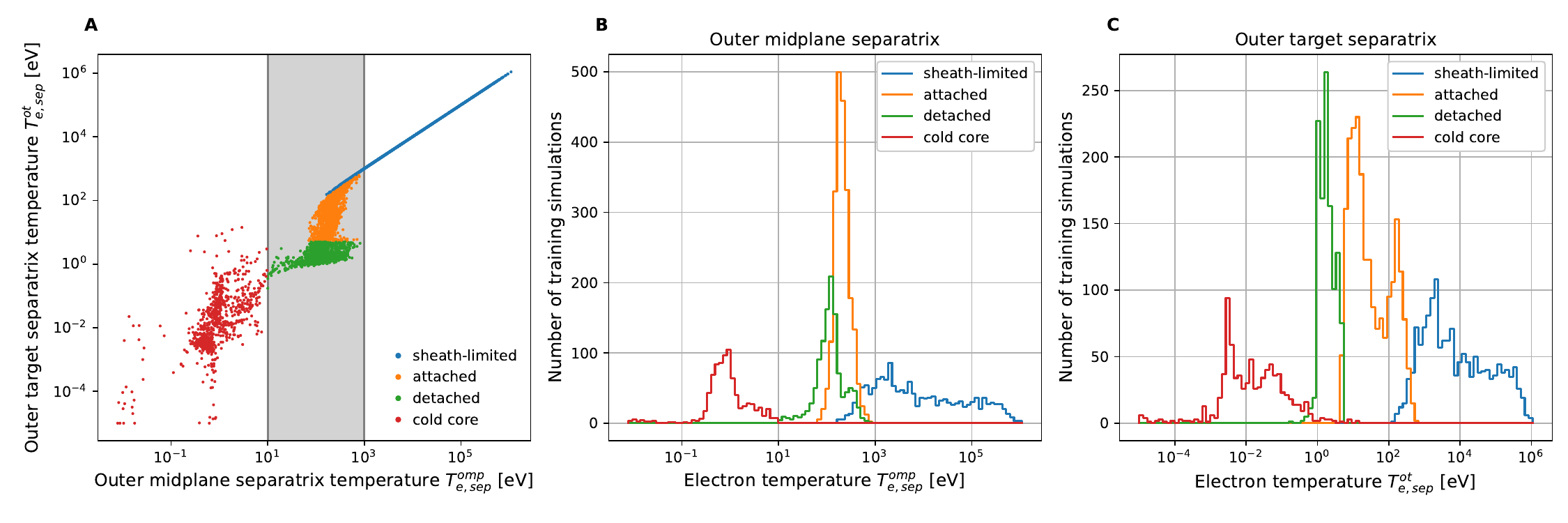}
    \caption{Electron temperature at the outer target separatrix against the electron temperature at the outer midplane separatrix in all training simulations (A). Histograms of the temperature at the outer midplane (B) and outer target (C) separatrix in all training simulations. The colors denote the different physical regimes of the simulations.}
    \label{fig:regimes}
\end{figure}

The primary dataset used in this study is an extension of the dataset developed in \cite{dasbach_towards_2023}. The reference SOLPS-ITER case for magnetic shape and wall geometry is based on a vertical target JET pulse 85423 ($2.5\,\mathrm{MA}$/$2.7\,\mathrm{T}$) that has been allowed to shrink in size (major radius $R$ and minor radius $a$) whilst keeping the aspect ratio $A=R/a$ and the plasma shear in terms of the safety factor profile $q_{95}=3.3$ constant.
In order to increase the capacity of the simulation dataset compared to \cite{dasbach_towards_2023} the total number of started simulations is increased to 8192 for training and 2048 for testing of models and the runtime is increased such that simulations were stopped only after ten seconds of simulated time if they did not converge to a steady-state or stable oscillations before. Around 30\% of the started simulations diverge and yield no results.
To facilitate the creation of such a large dataset the simulations use a fluid description for the neutral gas and less strict convergence criteria.
The deuterium ion and neutral gas inflow, nitrogen gas puff and the level cross-field transport coefficients (radial diffusion only) were varied, for a total of eight varied parameters. The scanned parameter ranges cover many machines (in terms of size: ASDEX-Upgrade, JET, ITER, DEMO) and are given in Table \ref{tab:params}.
For details concerning the simulation setup refer to \cite{dasbach_towards_2023}.
Besides covering different machine parameters the dataset is only useful if it also covers different physical regimes relevant for reactor operation.
This was already touched in \cite{dasbach_towards_2023} but since the classification into different regimes is used in the following sections a short analysis is presented here.
In the database always multiple parameters are varied at once.
This makes it much more difficult to classify the different divertor regimes using metrics such as the degree of detachment (DOD) \cite{loarte_plasma_1998}, which require a consistent density scan in a single scenario.
Instead the simulations are categorized by comparing the electron temperature directly outside the separatrix at the outer midplane $\teomp$ against the electron temperature outside of the separatrix at the outer (LFS) divertor target $\teot$ (Figure~\ref{fig:regimes}).
The simulations with very high temperatures show an almost perfect linear correlation between the target and upstream temperatures (Figure~\ref{fig:regimes}A).
Simulations with $\teot/\teomp \geq 80\%$ are categorized as sheath-limited because the only meaningful temperature gradient occurs in the plasma sheath at the divertor targets \cite{stangeby_plasma_2000}. These are 27\% of the simulations.
Simulations with larger temperature gradient and $\teot > 5\eV$ are called attached while similar simulations with $\teot \leq 5\eV$ are called detached. Of course with this broad distinction not all features of detachment (like momentum or particle losses) are necessarily achieved.
But some simulations with $\teot$ below $1\eV$ maintain upstream temperatures $\teomp > 100\eV$  (Figure~\ref{fig:regimes}A).
This demonstrates the high amount of heat reduction that takes place and suggest that low temperature effects like volume recombination become relevant \cite{stangeby_plasma_2000}.
With this distinction 38\%  of the simulations are attached and 20\% are detached.
Regardless of the temperature gradients all simulations with $\teomp < 10\eV$ are sorted into the cold core category.
Here the whole domain is cooled to less reactor relevant conditions and to levels the SOLPS-ITER code might not be particularly suited for \cite{coster_exploring_2017}. This concerns 27\% of the simulations.
Figure~\ref{fig:regimes}A depicts clearly that for $\teomp \in \left[100,1000\right]\eV$ moving from sheath-limited, over attached into detached conditions, the upstream temperatures remain largely at similar values while the target temperature decreases.
Only the transition into the cold core regime also leads to diminishing upstream temperatures.
Figures~\ref{fig:regimes}B~and~C depict the temperature statistics for the different regimes. For the sheath-limited cases the temperature distribution are identical at the outer midplane and outer target. The detached and attached cases have very sharply peaked distributions for $\teomp$ with maximums at $123\eV$ and $179\eV$, while the distributions at the target are much more diverse.
The developed surrogate models will have to deal both with these different statistics at different domain locations as well as with the overall large range of temperatures spanning several orders of magnitude.
The test set to evaluate the developed models is created in similar fashion as the training simulations and therefore follows similar statistics (not shown here).

\begin{table}
    \centering
	\begin{tabular}{|c|c|c|c|c|c|c|c|c|}
		\hline 
		& $R$ & $B$ & $\Dpuff$ & $\Npuff$ & $\Dcore$ & $\Pin$ & $D_{\perp}$ & $\chi_{\perp}$\tabularnewline
		\hline 
		\hline 
		min & $1$ & $1$ & $10^{20}$ & $10^{18}$ & $10^{19}$ & $10$ & $0.1$ & $0.1$\tabularnewline
		\hline 
		max & $10$ & $10$ & $10^{24}$ & $10^{23}$ & $10^{24}$ & $200$ & $2$ & $2$\tabularnewline
		\hline 
		units & $\m$ & $\T$ & $\atoms$ & $\atoms$ & $\atoms$ & $\MW$ & $\mmps$ & $\mmps$\tabularnewline
		\hline 
		scale & lin & lin & log & log & log & lin & lin & lin\tabularnewline
		\hline 
	\end{tabular}
	\caption{Overview of the parameters varied in the final dataset. The datapoints are distributed either uniformly over a linear scale (lin) or uniformly over a logarithmic scale (log) between minimum and maximum values.\label{tab:params}}
\end{table}

\newpage
\FloatBarrier
\section{Modelling the electron temperature}\label{sec:2D}
\begin{figure}
    \centering
    \includegraphics[width=\textwidth]{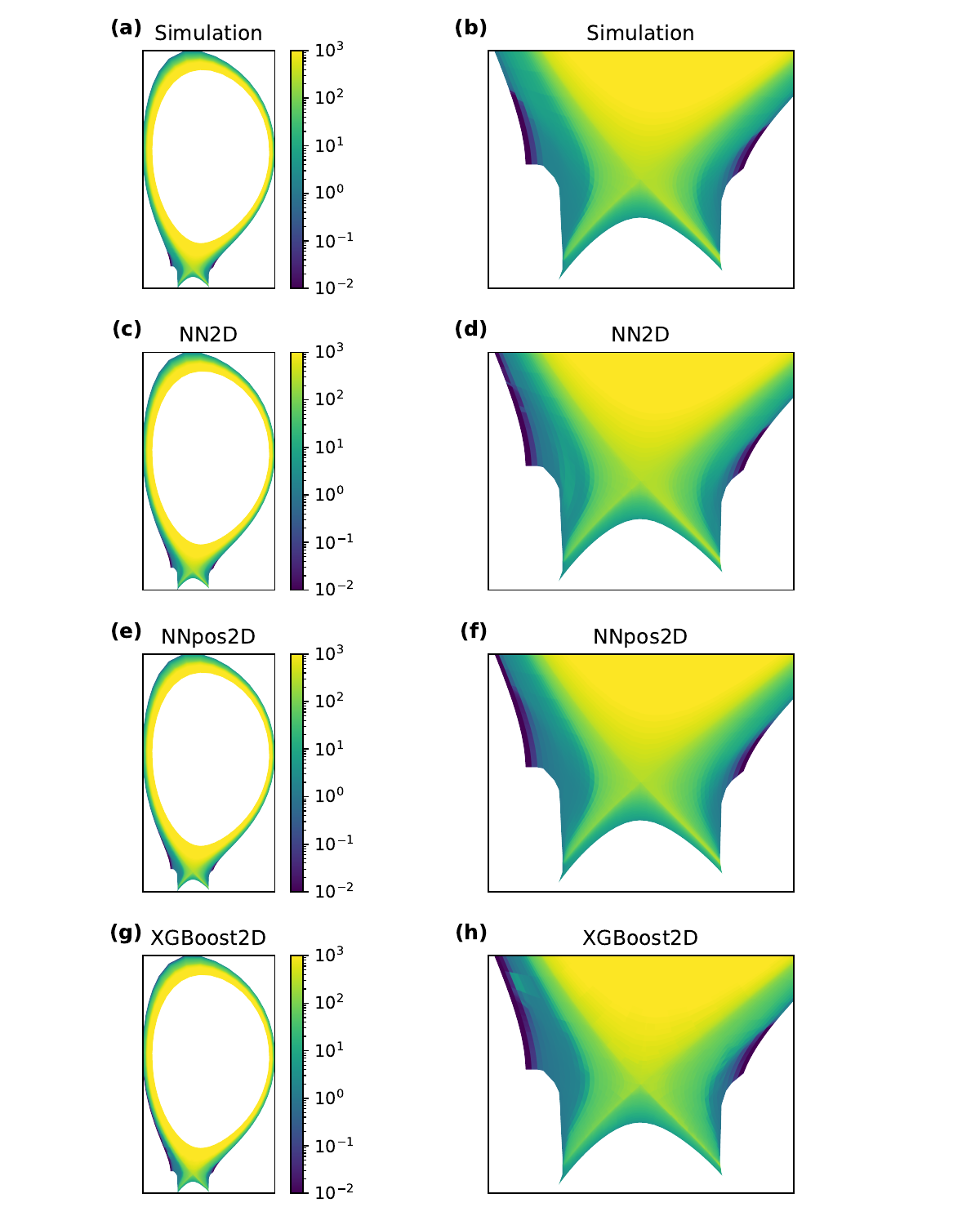}
    \caption{Comparison between the temperatures in the original SOLPS-ITER simulation and model predictions for one case from the test set. The right column shows the identical results as the left but zoomed to the divertor area, showing some artifacts in the profile (e.g. at the high-field side (HFS) region) from imperfect prediction.}
    \label{fig:domain2D}
\end{figure}

The developed SOLPS-ITER database is used in the following to train machine learning models.
To find a suitable model providing the full output as the original simulations, we split the problem into a series of steps until we arrive at the full model.
In the first step different model architectures are trained to predict the electron temperature in the whole 2D simulation domain.
Three different model architectures are compared.
The first model is a fully-connected neural network (NN2D) which receives the eight defining parameters of each simulation as input and is tasked to predict the corresponding electron temperatures $T_e$ in the whole simulation domain.
The simulation domain consists of 104x50 grid cells, thus the neural network has 5200 neurons in the last layer.
The network therefore tries to solve a regression task $\mathrm{NN2D}:\mathbb{R}^{8}\rightarrow\mathbb{R}^{104\times50}$.
The second model is also a fully-connected neural network (NNpos2D) but incorporates the 2D domain differently.
This model has only one neuron in the last layer and thus provides only a scalar $T_e$ output.
But in addition to the eight simulation parameters, the network receives the R and Z coordinates of each grid cell in the 2D domain as input. To predict the temperatures in the whole 2D domain for one simulation, this model thus has to perform 5200 independent predictions for each scalar temperature. This network tries to solve a regression task $\mathrm{NNpos2D}:\mathbb{R}^{10}\rightarrow\mathbb{R}^{1}$. Such a neural network architecture is similar to those used in Physics Informed Neural Networks (PINNs) \cite{wang_experts_2023}.
The third model is built using gradient boosted regression trees implemented in XGBoost (XGBoost2D).
Since XGBoost allows only for scalar outputs it employs a similar position dependent strategy and receives the R and Z coordinates of each grid cell as input. Therefore this model also tries to solve a regression task $\mathrm{XGBoost}:\mathbb{R}^{10}\rightarrow\mathbb{R}^{1}$.
Before training these models the simulation parameters and the temperatures are scaled to mean zero and variance one.
The exact procedure is summarized in \ref{app:HPO}.
All three models are trained in a random search with 5-fold cross validation to optimize the hyperparameters.
The optimal hyperparameters found are given in \ref{app:HPO}.
The following evaluates the final optimized models on the set of test simulations.
These test simulations were not used during the training and optimization process, thus the model errors are representative for the behaviour of the model when employed to make predictions for new scenarios.
Figure~\ref{fig:domain2D} shows the predicted temperatures by the various models compared to the simulation for one example case from the test set.
Overall the predictions of all models resemble the simulation data.
In this example the predicted temperatures by all models are too low near or above the inner target at the high-field side (HFS) compared to the original SOLPS-ITER simulation.
But the magnitude and location of such errors is different for each individual simulation in the test set.
\begin{figure}
    \centering
    \includegraphics[width=\textwidth]{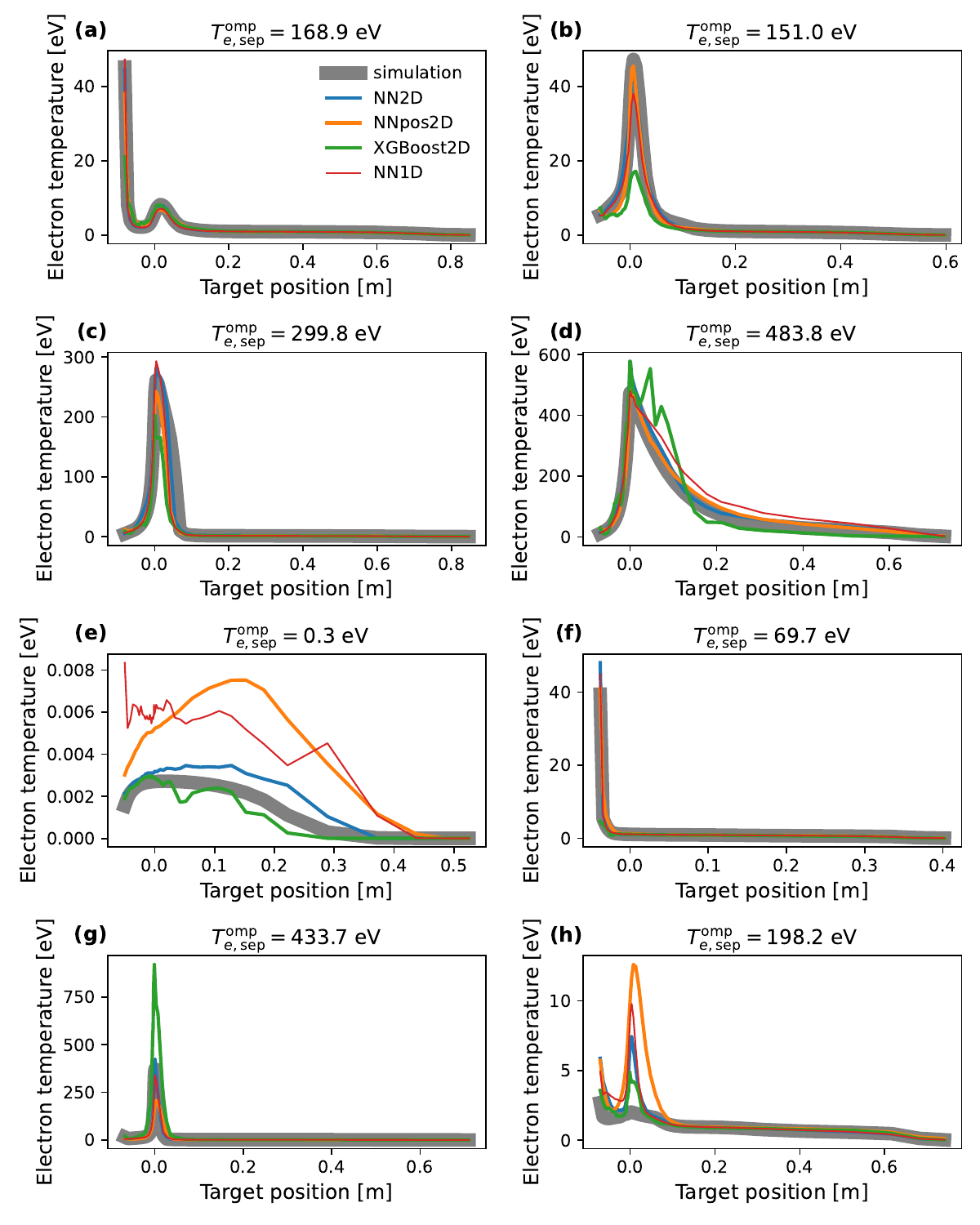}
    \caption{Comparison between the temperature profiles predicted by the various models at the outer target for eight SOLPS-ITER simulations from the test set. For reference the temperature at the outer midplane separatrix in each simulation is given. The different models are depicted by the colors: SOLPS-ITER simulation (grey), NN2D (blue), NNpos2D (orange), XGBoost2D (green), NN1D (red).}
    \label{fig:profiles2D}
\end{figure}
Further, Figure \ref{fig:profiles2D} shows predictions of the models just at the outer target for eight examples from the test set.
In those simulations with low target temperatures a temperature peak at the boundary of the outer target in the private flux region appears. This is an artifact in the underlying SOLPS-ITER simulations stemming from pumping model in the fluid neutral description.
Because the boundary of the simulation domain in the private flux region acts as pump (see \cite{dasbach_towards_2023}),
this is unnaturally close to the plasma, especially in simulations which are scaled down in size. A better tuning of the pumped particle fraction might remove this artifact.
Generally both the NN2D and NNpos2D models are able to produce accurate temperature profiles with both the correct shapes and overall correct magnitudes. However, in almost all examples there are some deviations between the model predictions and the ground truth simulations visible and in some cases like in Figure~\ref{fig:profiles2D}(h) the predictions are further off.
On these eight examples the XGBoost model is visibly less accurate than either of the neural network methods.
For comparison a fully connected neural network with predicts the temperatures only at the outer target was trained (NN1D, see \ref{app:HPO}). However, on these examples it does not appear more accurate than the models for the full 2D domain.
\begin{table}
    \centering
    \begin{tabular}{llrrrrrrr}
\toprule
 &  & $T_{e}$ - 2D & \multicolumn{5}{c}{$T^{ot}_{e}$ - 1D} & $T^{omp}_{e,sep}$ \\
 &  & all & all & \begin{tabular}{c}sheath-\\limited\end{tabular} & attached & detached & \begin{tabular}{c}cold\\core\end{tabular} & all \\
\midrule
\multirow[t]{2}{*}{NN2D} & abs & 3.3 & 0.97 & 8.7e+01 & 0.86 & 0.18 & 0.012 & 7.1 \\
 & rel & 0.065 & 0.12 & 0.082 & 0.11 & 0.12 & 0.68 & 0.043 \\
\cline{1-9}
\multirow[t]{2}{*}{NNpos2D} & abs & 3.8 & 1.6 & 8.6e+01 & 1.7 & 0.33 & 0.015 & 7.8 \\
 & rel & 0.068 & 0.19 & 0.092 & 0.21 & 0.21 & 0.68 & 0.041 \\
\cline{1-9}
\multirow[t]{2}{*}{XGBoost2D} & abs & 1.2e+01 & 2.2 & 3.8e+02 & 2.3 & 0.4 & 0.028 & 2.8e+01 \\
 & rel & 0.25 & 0.4 & 0.43 & 0.34 & 0.26 & 1.0 & 0.18 \\
\cline{1-9}
\multirow[t]{2}{*}{NN1D} & abs & 0.0 & 1.0 & 1e+02 & 0.92 & 0.19 & 0.015 & 0.0 \\
 & rel & 0.0 & 0.13 & 0.1 & 0.12 & 0.12 & 0.75 & 0.0 \\
\cline{1-9}
\bottomrule
\end{tabular}

    \caption{Median absolute and relative errors on the test set, calculated either over all grid points (2D), at the outer target (1D) or directly outside the separatrix at the outer midplane (OMP). The absolute errors are given $\eV$.}
    \label{tab:table2D}
\end{table}
For a quantitative comparison of the different models Table~\ref{tab:table2D} summarizes their median absolute and relative errors on the test set. Here the median errors are calculated either across all predictions for all locations in the 2D grid, across all locations at the outer target or at the outer midplane separatrix.
For evaluation of the test errors at the outer target, the test simulations are additionally split into the different regimes.
The NN2D model performs the best in all metrics.
It is obvious that the scale of the errors depends strongly on the absolute scales of the temperatures.
Therefore the absolute errors are highest at the outer midplane separatrix ($7.1\eV$), lower if calculated over the whole 2D domain ($3.3\eV$) and lowest at the low-field side (LFS) target ($0.97\eV$).
This highlights that any such metrics are useful to compare different models against each other on the same dataset, but not necessarily allow comparisons against models created on a different dataset.
Also across different regimes the absolute errors at the outer target are highest in the sheath-limited regime and smallest in the cold core regime. For the most relevant attached and detached regimes median errors of ($0.86\eV / 11\%$) and ($0.18\eV / 12\%$), respectively, are observed.
The NNpos2D network performs worse in all metrics.
Because the training times of the NNpos2D model are drastically higher fewer trials were conducted in the hyperparameter search compared to the NN2D model (see \ref{app:HPO}). Therefore, the decreased performance of the NNpos2D model might stem from a less optimal hyperparameter selection and with a more extensive optimization this architecture might still approach competitive levels of accuracy.
The XGBoost model obtains the worst scores and also requires hours to train, so is definitely the worst approach in this comparison.
Comparing the median errors of NN2D against the target specific model NN1D shows both to have similar levels of accuracy.
While the NN2D model obtains slightly better scores, the difference is not significant.
While the accuracy is far from perfect and would improve with more training data, one has to consider that even highly sophisticated SOLPS-ITER simulations rarely reproduce experimental results with deviations of less than $1\eV$.
Therefore, the discrepancy between the model and the fluid neutral simulations is likely much smaller than between the simulations and reality. Nevertheless, one has to consider that these metrics are statistics and there are no upper bounds for the errors the network models can make.
\\
To summarize the results of this Section: it is possible to develop surrogate models that predict the conditions for the whole scrape-off layer at once, without any compromise in terms of precision. It is therefore recommendable to only develop such full 2D models as they offer a wider range of applications without any major downsides compared to target specific models. The best model architecture found so far is a fully connected neural network with an output layer dimension identical to the computational grid.

\newpage
\FloatBarrier
\section{Model for several plasma properties}\label{sec:multi}
\begin{figure}
    \centering
    \includegraphics[width=0.8\textwidth]{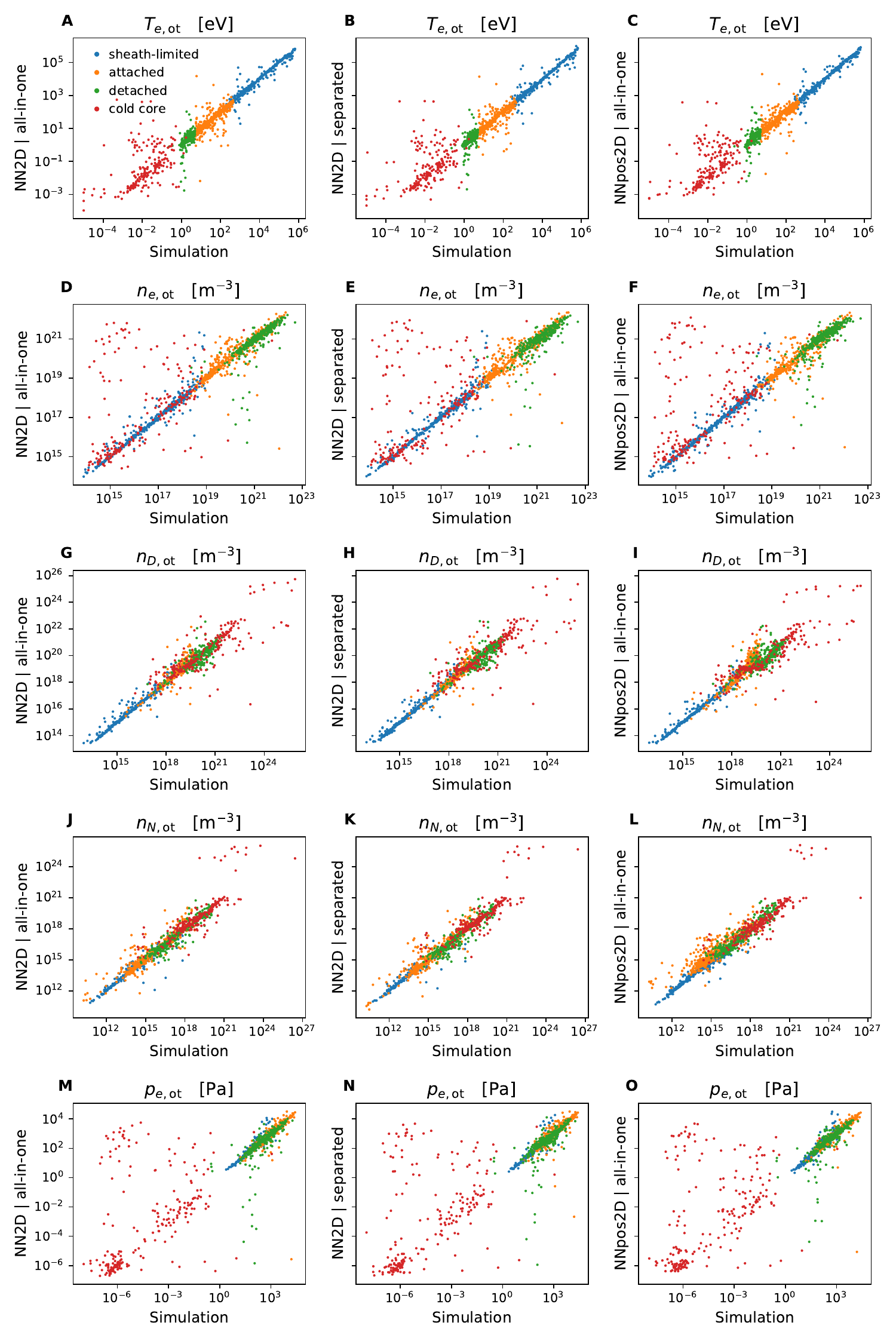}
    \caption{The model predictions at the outer target separatrix against the results in the simulations in the test set. The color denotes the regime of the simulation in the test set: sheath-limited (blue), attached (orange), detached (green), cold core (red).}
    \label{fig:multi-scatterOT}
\end{figure}

The previous analysis considers only the electron temperature.
Many more quantities are relevant for a full description of the SOL plasma.
In principle it should be possible to construct one large model predicting everything at once.
In practice it can be difficult to train such a large model and it might be easier to train independent models for each observable.
But independent models have no ability to profit from the correlations between different observables.
Which of these factors outweighs the other is specific to each problem.
To find a suitable strategy for the dataset here, both approaches are compared.\\
In addition to the NN2D network for the electron temperature $T_e$, three NN2D networks are trained independently to predict the electron density $n_e$, deuterium neutral gas density $n_D$ and nitrogen neutral gas density $n_N$.
For each of these networks a full hyperparameter search is conducted with the same procedure as in Section \ref{sec:2D}.
For comparison we setup two networks trained to predict all four quantities $T_e, n_e, n_D, n_N$ in the whole domain at once.
These architecture are also of the NN2D type with the exception that in one network the size of the output layer is increased to 4x5400 neurons (which is necessary to provide the output) while the second network has this increased size in the last two layers. Also for these two networks a similar hyperparameter search is conducted.
A variant of the NNpos2D architecture is also trained to predict all four observables simultaneously (NNpos2D-all-in-one). It shares the same structure as the network described in Section \ref{sec:2D} but features four scalar outputs instead of one. Its hyperparameters are fixed rather than optimized (see Table~\ref{tab:NN2D-search}).
For brevity the models are compared here only in terms of their predictions at the outer target separatrix. The qualitative conclusions remain the same, when the models are compared in the whole SOL or at the outer midplane \cite{dasbach_surrogate_2025}.
In addition to the predicted quantities, also the plasma pressure $p_e = n_e \cdot t_e$ derived from model predictions is analyzed.\\
Table~\ref{tab:multiOT} reports the median errors of the final optimized models on the test set.
In all quantities the separated models (NN2D-separated) are similarly accurate as the network combining all in one (NN2D-all-in-one). For electron temperature and neutral densities, the separated models achieve lower median errors, while for electron density and pressure the NN2D-all-in-one model is slightly ahead.
In comparison the combined network with the additional large layer before the output layer (NN2D-all-in-one-2) has in all cases slightly worse median errors than the NN2D-all-in-one model.
The NNpos2D-all-in-one model obtains similar scores as the NN2D-separated for the electron density and pressure, but seems worse the other quantities.
Interestingly, for each model the relative errors for all three densities are higher than for the temperature.
Figure~\ref{fig:multi-scatterOT} compares the outer target separatrix predictions of the models against the test simulations.
For most test cases all models predict values close to the simulation results.
For all models the strongest deviations from the optimal diagonal appear for simulations in the cold core regime.
No clear accuracy advantage is observed for either of the models.
The error metrics in Table~\ref{tab:multiOT} should not be over-interpreted. Especially the scores of the NN2D-separated and NN2D-all-in-one models are closely together. 
In terms of accuracy either approach is comparable.
In terms of extendability of the models, the NN2D-separated design has the big advantage that for each new observable an additional model can be trained and added without affecting the accuracy of all existing ones.
In the NN2D-all-in-one model, adding more observables would increase the necessary size of the output layer even more, which might impact the accuracy at some point.
Going forward we therefore employ separate networks for each observable.
The hyperparameter searches for each of the four tested observables showed that for each very similar network parameters achieved the best performance \cite{dasbach_surrogate_2025}. Going forward we therefore do not conduct a hyperparameter search but fix the network architectures to a good setup found in that scan (10 hidden layers with each 1000 selu neurons, learning rate 0.0001, L2 regularizer 0.0001, batch size 64).

\begin{table}
    \centering
    \begin{tabular}{llrrrrr}
\toprule
 &  & \multicolumn{5}{c}{OT} \\
 &  & $T_e$ & $n_e$ & $n_D$ & $n_N$ & $p_e$ \\
\midrule
\multirow[t]{2}{*}{NN2D-all-in-one} & abs & 3.1 & 9.4e+18 & 1e+18 & 5.5e+15 & 3.9e+01 \\
 & rel & 0.15 & 0.19 & 0.17 & 0.29 & 0.16 \\
\cline{1-7}
\multirow[t]{2}{*}{NN2D-all-in-one-2} & abs & 3.7 & 9.4e+18 & 1.3e+18 & 6.6e+15 & 4.4e+01 \\
 & rel & 0.19 & 0.25 & 0.25 & 0.38 & 0.19 \\
\cline{1-7}
\multirow[t]{2}{*}{NN2D-separated} & abs & 2.7 & 1.2e+19 & 9.5e+17 & 4.8e+15 & 5e+01 \\
 & rel & 0.14 & 0.23 & 0.18 & 0.27 & 0.2 \\
\cline{1-7}
\multirow[t]{2}{*}{NNpos2D-all-in-one} & abs & 3.4 & 1e+19 & 2.3e+18 & 1.2e+16 & 4.7e+01 \\
 & rel & 0.16 & 0.23 & 0.35 & 0.55 & 0.2 \\
\cline{1-7}
\bottomrule
\end{tabular}

    \caption{Median absolute and relative errors on the test set of the model predictions at the outer target separatrix. The absolute errors are given in the units $\left[T_e\right]=\eV$, $\left[n_e\right]=\left[n_D\right]=\left[n_N\right]=\mathrm{m}^{-3}$,$\left[p_e\right]=\mathrm{Pa}$.}
    \label{tab:multiOT}
\end{table}

\newpage
\FloatBarrier
\section{SOLPS in the Loop}\label{sec:loop}
\begin{figure}
    \centering
    \includegraphics[width=\textwidth]{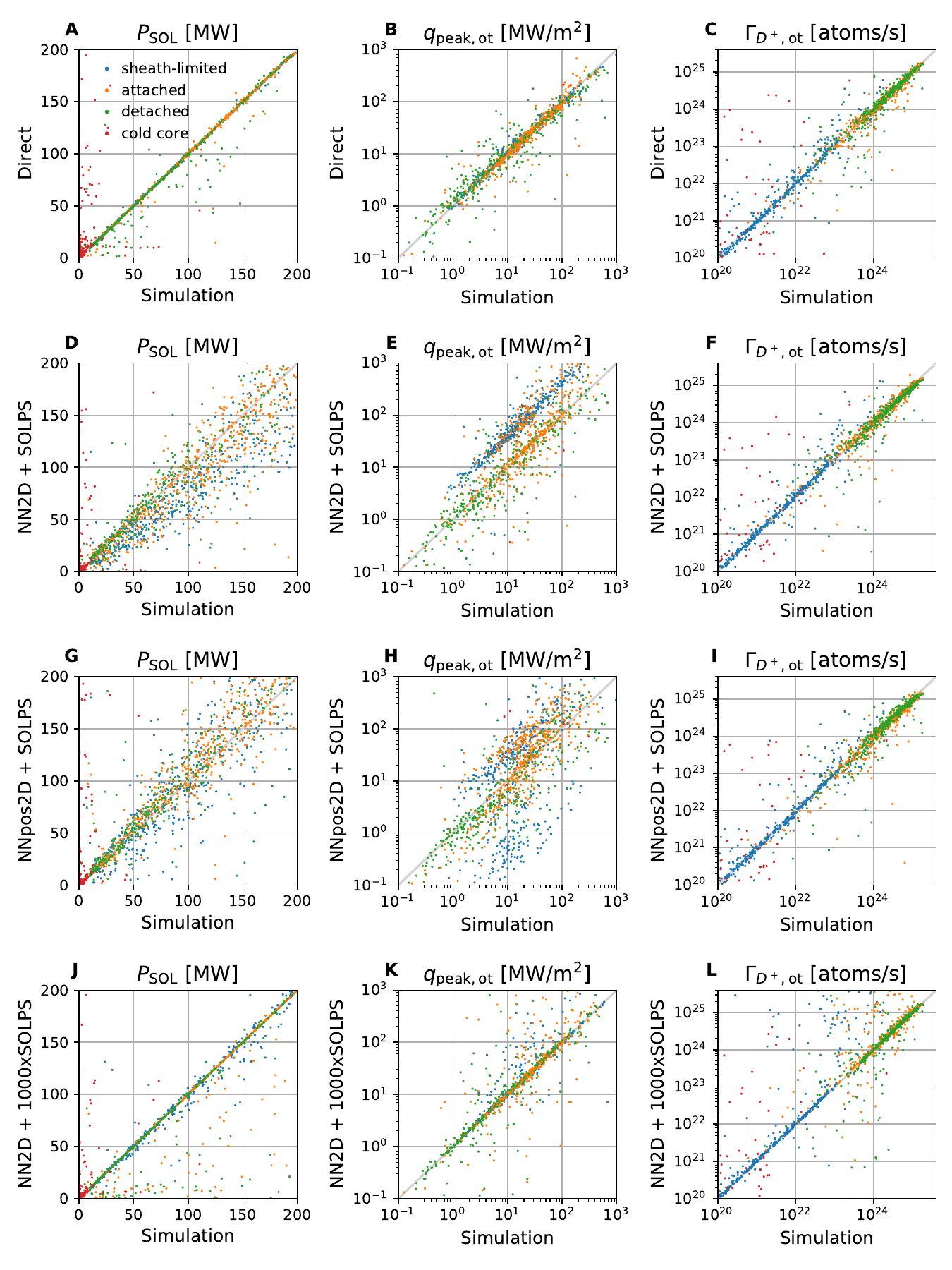}
    \caption{The model predictions for the three QoI against the results in the simulations in the test set. The color denotes the regime of the simulation in the test set. The lower axis limits in B,E,H,K are adjusted to highlight the systematic shift in the predictions. Therefore the cold core cases with smaller heat fluxes are not seen.}
    \label{fig:aux1}
\end{figure}

The list of all quantities of interest (QoI) in the SOL can become quite extensive as different applications require different properties.
Heat and particle fluxes at the divertor are relevant for determining safe exhaust regimes, coupling to pedestal models requires densities and temperatures at the separatrix, comparison with experiment might require radiation profiles or line integrated quantities, etc.
For each QoI a separate model can be trained, but this seem impractical as for niche use cases, users might be required to create their own custom models.
An alternative is to predict all independent state variables of SOLPS-ITER simulations, from which any possible QoI can be derived.
This section compares these two approaches on three example QoIs: power crossing the separatrix $P_{SOL}$ , the peak heat flux at the outer divertor target $q_{peak,ot}$ and the integrated deuterium ion flux at the outer divertor target $\Gamma_{D_+,ot}$.
For each of the three QoI a network with scalar output is trained and hyperparameters optimized.
Depending on the exact settings of the SOLPS-ITER simulations, the complete list of state variables can vary slightly.
Here the state variables are the electron temperature $T_e$, ion temperature $T_i$, densities of all ion and neutral particles ($n_D,n_{D+},n_N,n_{N+},...,n_{N7+}$) and the parallel velocities of all ions and neutral particles ($u_D,u_{D+},u_N,u_{N+},...,u_{N7+}$).
The electron density follows from quasineutrality and the potential equation is not solved but a potential of $3.1T_e/e$ is assumed.
For each of these 22 variables a separate NN2D model with fixed hyperparameters is trained.
Additionally, a network based on the NN2pos2D architecture is trained with the same hyperparameters as before but with 22 scalar outputs (one for each state variable).
To derive the three QoI from the network predictions it is not necessary to implement their formulas again but instead SOLPS-ITER can be used.
To do so the predicted state variables are written into SOLPS-ITER input files (b2fstati).
This allows us to compute all dependent quantities in one step with a guarantee that the same formulas are used as in the original simulations.
But care needs to be taken to make sure that the simulation really recomputes all dependent quantities from the state variables (see \cite{dasbach_surrogate_2025} Chapter 13).
\begin{table}
	\centering
	\begin{tabular}{lllrrrrr}
\toprule
          &                            &     &     all &  sheath-limited &  attached &  detached &  cold core \\
\midrule
\multirow{6}{*}{\rotatebox[origin=c]{90}{Direct}} & \multirow{2}{*}{$P_{\mathrm{SOL}}$} & abs &    0.51 &            0.38 &      0.51 &      0.61 &        1.3 \\
          &                            & rel &  0.0076 &          0.0048 &    0.0057 &    0.0096 &       0.46 \\
\cline{2-8}
          & \multirow{2}{*}{$q_{\mathrm{peak,ot}}$} & abs &    0.98 &            0.71 &       1.7 &       1.9 &      5e-05 \\
          &                            & rel &    0.13 &           0.055 &     0.092 &      0.34 &       0.98 \\
\cline{2-8}
          & \multirow{2}{*}{$\Gamma_{D^+,\mathrm{ot}}$} & abs & 6.5e+22 &         5.4e+20 &   2.1e+23 &   2.5e+23 &    3.6e+19 \\
          &                            & rel &    0.15 &            0.15 &     0.085 &      0.15 &        6.7 \\
\cline{1-8}
\cline{2-8}
\multirow{6}{*}{\rotatebox[origin=c]{90}{NN2D}} & \multirow{2}{*}{$P_{\mathrm{SOL}}$} & abs & 1.1e+01 &         1.7e+01 &   1.1e+01 &       9.3 &       0.77 \\
          &                            & rel &    0.19 &            0.24 &      0.13 &      0.16 &       0.72 \\
\cline{2-8}
          & \multirow{2}{*}{$q_{\mathrm{peak,ot}}$} & abs & 1.1e+01 &         4.4e+01 &       8.7 &       2.5 &    0.00076 \\
          &                            & rel &     1.7 &             3.1 &      0.43 &      0.41 &    1.5e+02 \\
\cline{2-8}
          & \multirow{2}{*}{$\Gamma_{D^+,\mathrm{ot}}$} & abs & 7.1e+22 &         4.4e+20 &   1.9e+23 &   2.5e+23 &    2.3e+19 \\
          &                            & rel &    0.15 &            0.13 &     0.089 &      0.16 &        1.0 \\
\cline{1-8}
\cline{2-8}
\multirow{6}{*}{\rotatebox[origin=c]{90}{NNpos2D}} & \multirow{2}{*}{$P_{\mathrm{SOL}}$} & abs &     9.4 &         1.4e+01 &     1e+01 &       7.8 &       0.86 \\
          &                            & rel &    0.15 &            0.18 &      0.12 &      0.13 &       0.49 \\
\cline{2-8}
          & \multirow{2}{*}{$q_{\mathrm{peak,ot}}$} & abs &     9.8 &         1.6e+01 &   1.7e+01 &       2.7 &      0.025 \\
          &                            & rel &    0.91 &            0.98 &       0.6 &      0.49 &    3.4e+03 \\
\cline{2-8}
          & \multirow{2}{*}{$\Gamma_{D^+,\mathrm{ot}}$} & abs &   7e+22 &           5e+20 &     3e+23 &   3.1e+23 &    2.1e+19 \\
          &                            & rel &    0.17 &            0.12 &      0.13 &      0.18 &        1.0 \\
\cline{1-8}
\cline{2-8}
\multirow{6}{*}{\rotatebox[origin=c]{90}{NN2D+1000}} & \multirow{2}{*}{$P_{\mathrm{SOL}}$} & abs &    0.39 &             1.4 &      0.14 &      0.19 &       0.53 \\
          &                            & rel &   0.011 &           0.023 &    0.0016 &    0.0033 &       0.24 \\
\cline{2-8}
          & \multirow{2}{*}{$q_{\mathrm{peak,ot}}$} & abs &    0.79 &            0.75 &       1.4 &       2.3 &    3.6e-05 \\
          &                            & rel &    0.12 &           0.046 &     0.074 &      0.23 &       0.99 \\
\cline{2-8}
          & \multirow{2}{*}{$\Gamma_{D^+,\mathrm{ot}}$} & abs &   7e+22 &         3.2e+20 &   2.1e+23 &     2e+23 &      7e+18 \\
          &                            & rel &    0.12 &            0.08 &     0.084 &      0.12 &        1.3 \\
\bottomrule
\end{tabular}

	\caption{Median absolute and relative errors on the test set of the three model predicted QoI. The errors are evaluated either on the whole test set or only those simulations in certain regimes (as defined in Section \ref{sec:dataset}). The absolute errors are given in the units $\left[P_{\mathrm{SOL}}\right]=\MW$, $\left[q_{\mathrm{peak,OT}}\right]=\mathrm{MW}/\mathrm{m}^2$, $\left[\Gamma_{D^+,\mathrm{OT}}\right]=\atoms$.}
	\label{tab:aux}
\end{table}
Table \ref{tab:aux} compares the median errors on the test set of the QoI predicted by the various methods. All three approaches (Direct, NN2D+SOLPS, NNpos2D+SOLPS) yield similar errors for $\Gamma_{D_+,ot}$ across all regimes.
The direct approach predicts $P_{SOL}$ with high accuracy except in cold core conditions.
In contrast, NN2D+SOLPS and NNpos2D+SOLPS both yield median errors larger than 10\% for $P_{SOL}$ across all regimes.
For the peak heat-flux density at the outer target $q_{peak,ot}$ the direct approach performs best in the sheath-limited regime, while NN2D+SOLPS and NNpos2D+SOLPS struggle there, reaching median errors up to 310\%.
In the attached regime the direct model is still far more accurate and only in the detached regime both other approaches catch up, while still being slightly less accurate.
The large relative errors in the cold core cases stem from near-zero absolute values.
Figure \ref{fig:aux1} gives further insight into the errors of the NN+SOLPS scheme.
Here the predictions are compared against the ground-truth values in the test simulations, such that for a perfect prediction all points lie on the diagonal.
Predicting the QoI directly shows to be very accurate for all three quantities (Figure \ref{fig:aux1}A-C).
The NN2D+SOLPS approach appears similarly accurate for the ion flux towards the outer target (Figure \ref{fig:aux1}F) but less accurate for $P_{SOL}$ (Figure \ref{fig:aux1}D).
For the peak target heat flux predictions the results are more complicated (Figure \ref{fig:aux1}E).
In "detached" conditions the NN2D+SOLPS approach is accurate.
But in all sheath-limited and some of the attached cases the predictions show a systematic shift towards higher values.
This explains why the median errors of the NN2D+SOLPS approach are comparable to the direct predictions for detached cases but much worse in sheath-limited and attached regimes (Table \ref{tab:aux}).
The NNpos2D+SOLPS approach shows a similar behaviour. High accuracy for the target particle flux (Figure \ref{fig:aux1}I), slightly worse accuracy for $P_{SOL}$ (Figure \ref{fig:aux1}G) and a systematic deviation for sheath-limited and attached cases in $q_{peak,ot}$ (Figure \ref{fig:aux1}H).
Although here also the predictions for detached cases show a systematic shift to lower values (Figure \ref{fig:aux1}H).
The cause of these errors can be found in the physics of the heatflux calculations.
The conductive heatflux follows a relation of $q_{cond} \propto T^{5/2} \partial T / \partial x$.
In the sheath-limited (and the high temperature attached) regime the temperatures are almost isothermal in a fluxtube but with absolute values of hundreds to thousands of eV (Figure \ref{fig:regimes}A). Due to this even microscopic temperature gradients elicit a finite heatflux towards the divertor targets.
Because the simulations are almost isothermal the temperature differences between neighbouring locations in the simulations ($\delta T_e < 0.001 \eV$) are a lot smaller than the accuracy of the networks.
The temperature gradients derived from the network predictions are thus much larger in magnitude and point in random directions, because there is almost no spatial correlation to the neural network errors.
Since $q_{peak,ot}$ is defined as the highest heatflux arriving at any location on the outer divertor target, this noise in the heatflux profiles will yield a systematic over prediction.
Vice versa in the detached simulations the temperature gradients are larger and thus the local temperature differences are larger than the errors of the networks.
Thus the heatfluxes derived from the predictions in these regimes are much closer to the groundtruth.
The fact that $P_{SOL}$ is less effected by this than $q_{peak,ot}$ has two reasons.
Mainly because the radial temperature differences across the separatrix are larger and secondly because $P_{SOL}$ is an integrated quantity which allows some level of noise to average out.
A possible remedy to these errors could lie in moving to more complex neural network architectures and training procedures, which enforce physics-informed spatial correlation.
However, this is not straight forward and it can be difficult finding the right balance between predicting the most accurate absolute values and maintaining spatial correlations.
Instead SOLPS-ITER can be used to smooth out the mistakes in the temperature gradients.
The workflow stays exactly the same but instead of computing only the depending quantities, the simulation is iterated from the network prediction for several timesteps.
Table \ref{tab:aux} and Figure \ref{fig:aux1}J, K, L depict the results after the NN2D predictions have been iterated for 1000 steps.
The errors for all three QoIs have improved drastically compared to the plain NN2D+SOLPS approach.
Specifically, for $q_{peak,ot}$ the systematic deviations have decreased and the accuracy surpasses even the direct network predictions.
Of course this approach is much slower than purely evaluating a neural network, but since the original simulations were each run between $10^5-10^6$ steps still yields a speedup of factor 100-1000 compared to running a simulation from scratch.
More work is needed to systematically study what speedups are realistic using this approach for different simulations settings and if the numerical simulation parameters can be specifically optimized for it.

\newpage
\FloatBarrier
\section{Model validation}\label{sec:physics}
\begin{figure}
	\centering
	\includegraphics[width=\textwidth]{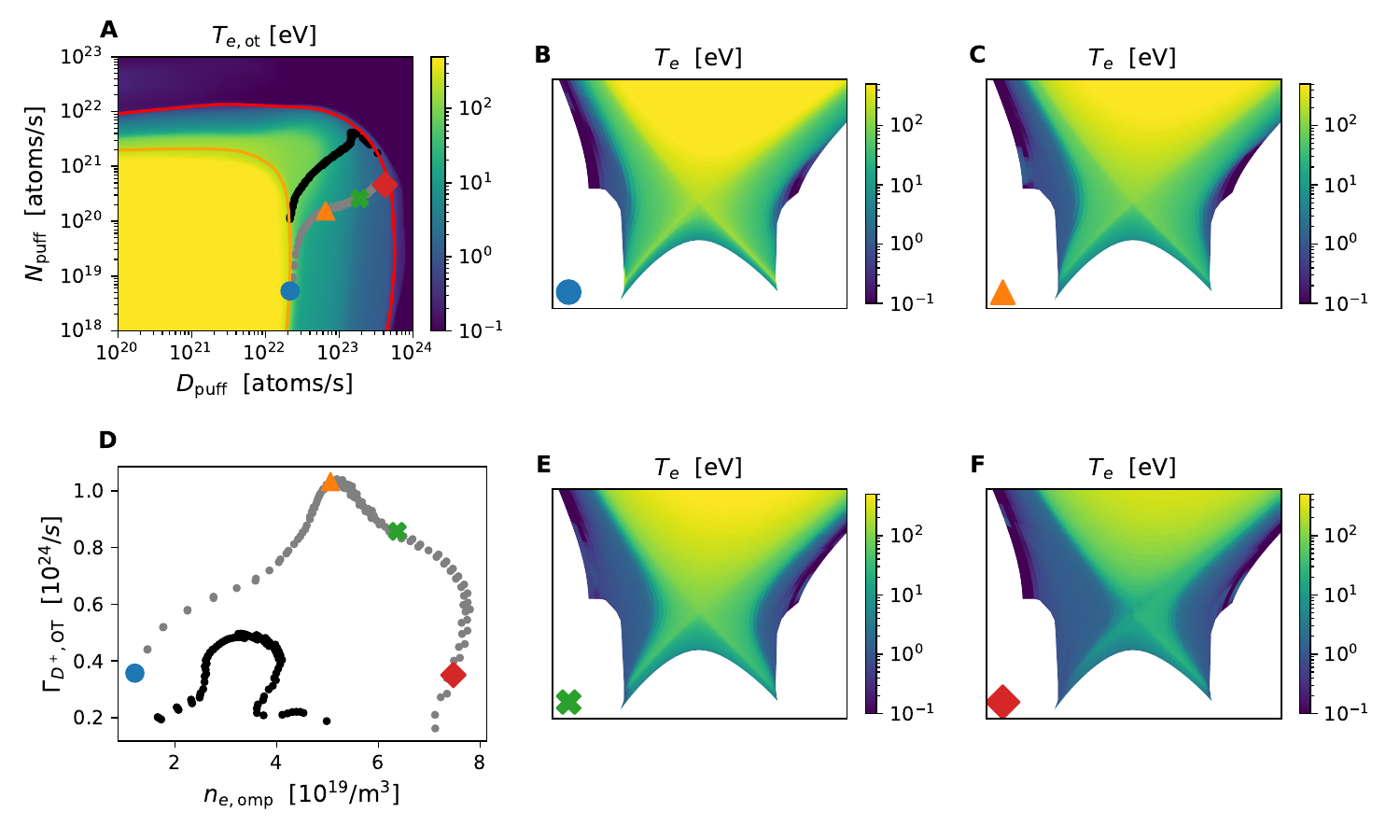}
	\caption{Predictions of the surrogate model for a gas puff scan with \textbf{JET} parameters. A: The outer target temperature at the separatrix $T_{e,\ot}$. The orange and red lines highlight the transition into the sheath-limited and cold core regimes, respectively. The grey and black dots show all scenarios with nitrogen concentrations $c_{N,\mathrm{div}}=0.1\%$ and $c_{N,\mathrm{div}}=1\%$, respectively. D: The integrated deuterium ion flux $\Gamma_{D^+,\ot}$ at the outer target as function of the electron density at the  outer midplane separatrix $n_{e,\omp}$ for the scenarios marked by the grey and black curves in A. (B,C,E,F): The model predicted electron temperatures for four scenarios marked by colored symbols in A and D.}
	\label{fig:physics-scan}
\end{figure}

\begin{figure}
    \centering
    \includegraphics[width=\textwidth]{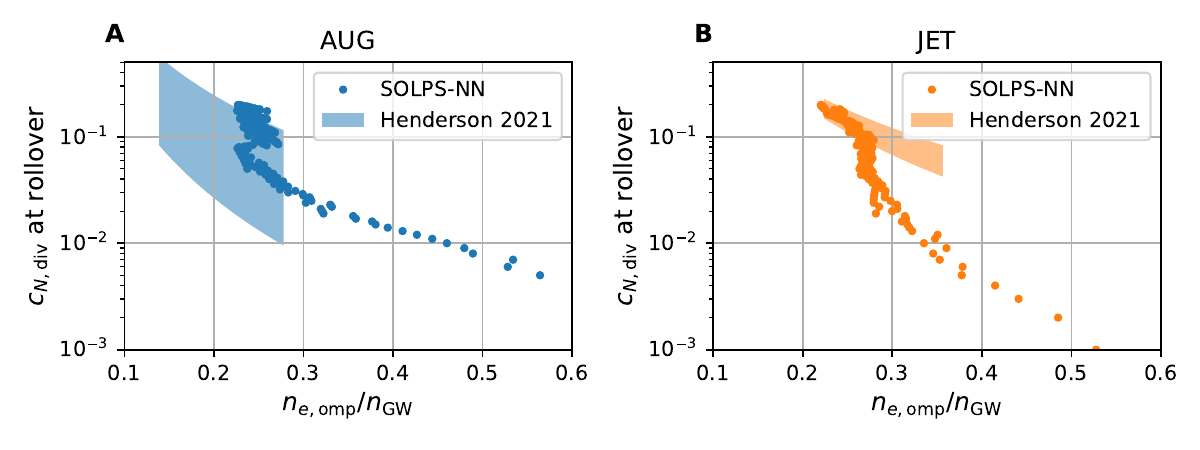}
    \caption{Electron densities at the outer midplane separatrix and divertor nitrogen concentrations at particle flux rollover for ASDEX-Upgrade (left) and JET (right). The rollover is defined as the point of maximum deuterium ion flux to the outer target for fixed nitrogen concentration. For comparison across tokamaks the densities (horizontal axes) are normalized with the Greenwald densities, c.f. Table \ref{tab:goldston}. The experimental fits are given by Equation (3) and (6) in \cite{henderson_parameter_2021}.}
    \label{fig:physics-scaling}
\end{figure}

In the previous sections it has been verified that the neural networks reproduce the underlying simulations.
But since the fidelity of these simulations is reduced, the developed models are only useful if they show meaningful physical behavior.
Due to the assumptions underlying the model (fixed geometry during size scaling, fluid neutrals, flat transport, etc.)
we do not expect to reproduce the results of any experimental tokamak perfectly.
But as long as the trends are realistic, these models might be sufficient to be used in very fast first assessment of new tokamak design points.
A central question are divertor design the impurity concentration and plasma density required for detachment \cite{pacher_impurity_2009,kallenbach_impurity_2013,wiesen_plasma_2017,putterich_impurity_2015} and several scaling relations have been proposed to predict these requirements \cite{goldston_new_2017,kallenbach_analytical_2016,reinke_heat_2017}.
A full validation of this model is beyond the scope of this paper and should be pursued in subsequent publications.
Therefore this sections gives only a very brief summary of the more extensive analysis shown in \cite{dasbach_surrogate_2025} with the aim to assess whether the models can predict the density and impurity concentrations necessary to enter detachment.
In contrast to the analysis in \cite{dasbach_surrogate_2025} slightly different parameters are used to represent the tokamaks and the nitrogen concentration is measured in the divertor instead of upstream.
\\
The models from Section~\ref{sec:loop} (directly predicting 2D plasma and $\Gamma_{D_+,\mathrm{OT}}$) are used here to run a parameter scan for JET.
The parameters which remain fixed during this scan are given in Table~\ref{tab:goldston}.
The deuterium and nitrogen puff rates are varied logarithmically over 200 steps, resulting in 40000 predicted conditions.
Figure~\ref{fig:physics-scan} shows the results of this scan.
The plasma temperatures at the target are extremely high for low gas puff levels, but decrease with increasing gas puffs (Figure~\ref{fig:physics-scan}A).
The orange and red contours in the figure mark transitions into sheath-limited and cold core regimes (as in Section~\ref{sec:dataset}).
Only a narrow range of gas puffs supports attached or detached conditions, while too excessive puffing will cause the plasma inside the confined field lines to cool down significantly.
The attached/detached band spans diagonally from low nitrogen/high deuterium (bottom right) to the inverse (top left), but only the lower part has desirable levels of impurities for reactor conditions (e.g. ITER aims at $c_\mathrm{Ne} < 1\%$ \cite{pacher_impurity_2009,pacher_impurity_2015}).
The divertor nitrogen concentration is calculated as $c_\mathrm{N,div} = \int(n_\mathrm{N} + n_\mathrm{N^+} + \dots + n_\mathrm{N^{7+}}) \mathrm{dV} / \int n_\mathrm{e}\mathrm{dV}$ with volume average over the divertor area depicted in Figure~\ref{fig:impurity-zone}.
In Figure~\ref{fig:physics-scan}A attached and detached scenarios with divertor $c_\mathrm{N,div}$ of 0.1\% (grey) and 1\% (black) are highlighted.
A fixed nitrogen concentration appears as roughly diagonal contour, so at a constant ratio between the nitrogen and deuterium gas puff.
Movement along these lines reflects changes in plasma density and neutral content.
Figure~\ref{fig:physics-scan}D shows the deuterium ion flux towards the outer target for the two highlighted traces ($c_\mathrm{N,div}$=0.1\% and $c_\mathrm{N,div}$=1\%).
Both levels of nitrogen concentration experience a rollover of the particle flux with increasing density.
As expected the rollover occurs at lower plasma density for higher nitrogen concentration.
At a point beyond the rollover the upstream density stagnates and eventually begins to decline with increasing gas puffs.
A similar behavior is reported in other SOLPS studies for ITER \cite{pacher_impurity_2015,moulton_comparison_2021}.
Also this maximum reachable density decreases with increasing nitrogen concentration.
Higher densities might be reached when fueling is conducted through the core instead of gas puffing.
The decreasing upstream density coincides with approaching the cold core regime, where the confined plasma gets cooled strongly.
This obfuscates the correlations to some extent. In particular, the black curve in Figure~\ref{fig:physics-scan}A is no longer diagonal, and this deviation is responsible for the slight increase in upstream density at low particle flux after the rollover and density stagnation observed in Figure~\ref{fig:physics-scan}D. As the nitrogen concentration is not perfectly aligned with a constant puff ratio, changes in fueling simultaneously modify density, neutral content, and radiation distribution, leading to a more complex trajectory in parameter space.
Figures~\ref{fig:physics-scan}B, C, E, and F depict the spatial electron temperature distributions predicted by the model for four representative conditions along the $c_\mathrm{N,div}=0.1\%$ trace. These panels illustrate the progressive cooling of the scrape-off layer (SOL) as the gas puff is increased. In the model detachment initiates at the outermost radial positions and propagates inwards. This behavior is in contrast to the expected pattern, where detachment typically starts close to the strike point. The deviation is most likely related to the simplified fluid neutral treatment and the location of the gas puff at the outer boundary of the computational domain.
As a result, the neutral density builds up first in the outer SOL, leading to enhanced radiative cooling in this region.
In Figure~\ref{fig:physics-scan}F the SOL is almost entirely cooled down.
However, the spatial pattern remains different from experimental observations.
Instead of a neutral cloud localized in the divertor pushing the plasma upstream, the neutrals push inward all around the confined plasma.
A further increase in gas puff beyond the conditions shown in Figure~\ref{fig:physics-scan}F would eventually lead to strong cooling of the confined plasma, effectively pushing the plasma boundary back into the closed field line region and driving the system into the cold core regime.
Based on the gas puff scan shown in Figure~\ref{fig:physics-scan}, the plasma density required to reach detachment can now be extracted for various nitrogen concentrations.
For simplicity, the onset of detachment at a given nitrogen concentration is defined here as the point of maximum deuterium ion flux to the outer target, independent of whether specific heat flux or temperature criteria are fulfilled.
The same procedure is repeated using parameters representative of ASDEX Upgrade (AUG) as listed in Table~\ref{tab:goldston}. The resulting densities at detachment onset are compared with experimental fits reported in \cite{henderson_parameter_2021}. For AUG, the fit reads
$c_{N,\mathrm{div}} = 21.9 \cdot \left(P_\mathrm{div,outer}/\mathrm{MW}\right)^{1.24\pm0.45}
\left(n_{e,\mathrm{omp}}/10^{19}\mathrm{m}^{-3}\right)^{-2.71\pm0.41}$, valid in the range $n_{e,\mathrm{omp}} = (2.0–4.0)\cdot10^{19},\mathrm{m}^{-3}$.
Following \cite{henderson_parameter_2021}, we approximate $P_\mathrm{div,outer}=\Pin/2.37$.
For JET, the corresponding experimental fit is
$c_{N,\mathrm{div}} = 1.245\left(n_{e,\mathrm{omp}}/10^{19}\mathrm{m}^{-3}\right)^{-2.43}$, valid within $n_{e,\mathrm{omp}} = (2.2–3.5)\cdot10^{19}\mathrm{m}^{-3}$. The comparison between surrogate predictions and experimental scalings is shown in Figure~\ref{fig:physics-scaling}.
For both AUG and JET the SOLPS-NN surrogate predicts the expected qualitative trend that higher plasma densities are required to reach detachment at lower nitrogen concentrations. In the case of AUG, the surrogate predictions agree with the experimental scaling aorund 20–30\% of the Greenwald density. However, a saturation behavior is observed: for nitrogen concentrations above about 10\%, the density at rollover no longer decreases significantly but remains close to 25\% of the Greenwald density. This behavior is unlikely to be physical. Rather than exhibiting a pronounced rollover, the system may already be detached at the lowest density above the sheath-limited regime. At higher densities, the surrogate follows a smoother and more continuous trajectory.
For JET the surrogate predictions similarly match the experimental scaling in the range of roughly 20–28\% of the Greenwald density. But at higher densities the predicted impurity concentration drops rapidly and becomes nearly an order of magnitude smaller than the experimental values around 35\% of the Greenwald density. In this regime the network is therefore overly optimistic, predicting access to detachment at substantially lower impurity concentrations than observed experimentally.
In summary, despite the reduced fidelity of the underlying training simulations, the surrogate models reproduce the experimentally observed access conditions for detachment with surprisingly good accuracy at moderate densities. At the same time, the unphysical spatial behavior seen in Figure~\ref{fig:physics-scan} and the deviations at higher densities demonstrate that these results should not be over-interpreted. And for other quantities of interest the deviations might be significantly larger.
Nevertheless, the models do not yield qualitatively unreasonable predictions, which supports confidence in the general modeling approach. This suggests that the same surrogate-building procedure, when applied to higher-fidelity simulations, could provide more reliable predictive tools.
It also indicates that cross-machine surrogates not tailored to a specific device may still achieve sufficient accuracy for broad scoping studies and are at the very least comparable to simple experimental scaling laws.

\newpage
\FloatBarrier
\section{Transfer Learning on ITER data}\label{sec:transfer}
\begin{figure}
    \centering
    \includegraphics[width=\textwidth]{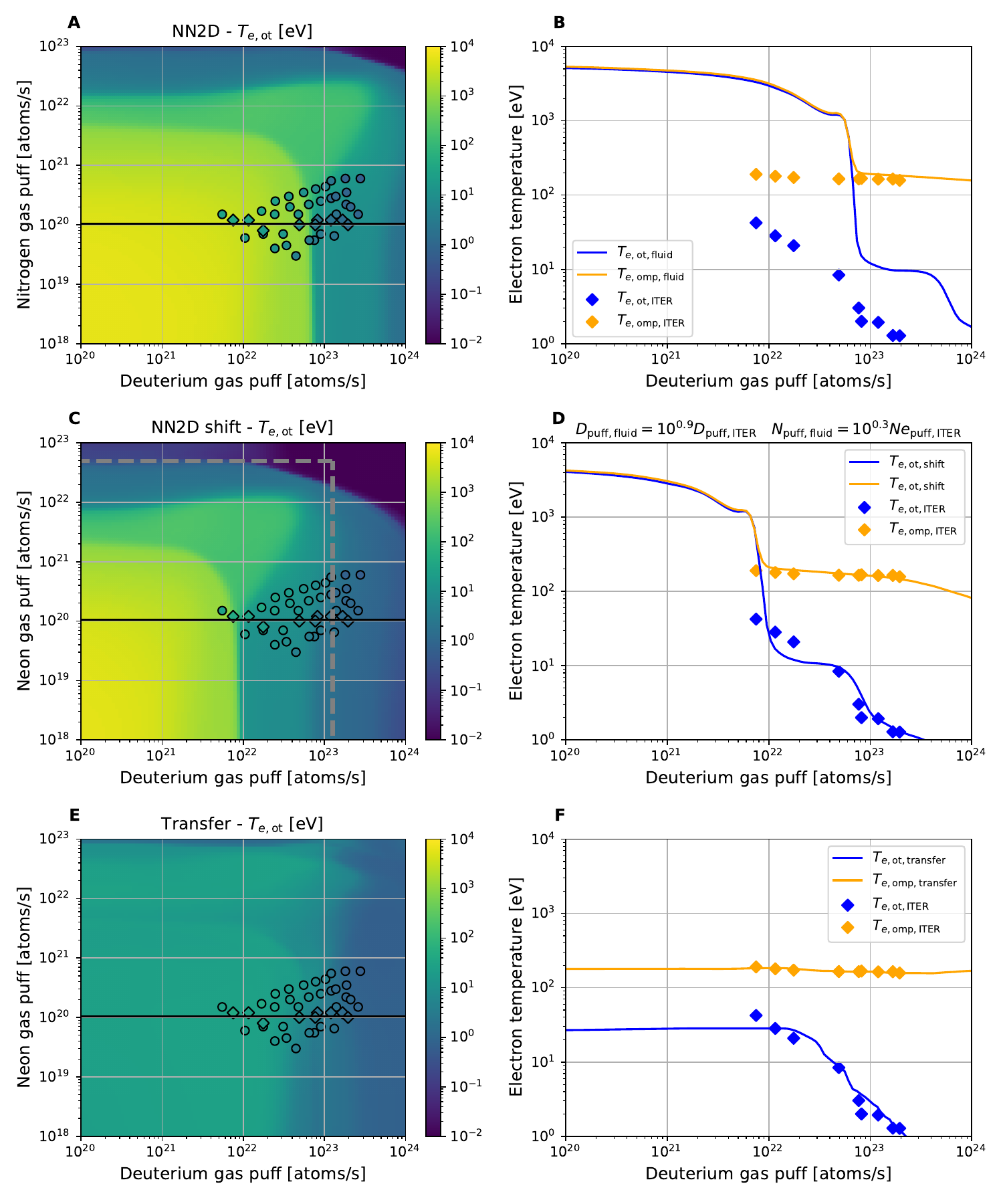}
    \caption{Temperatures predicted by the NN2D fluid neutral model (A,B), the model with rescaled gas puff (C,D) and the transfer learned model (E,F) in comparison to the temperatures in the ITER simulation (markers on the left) with $100\MW$ input power. The color inside the markers on the left corresponds to the temperature in the ITER simulations. The dashed grey lines in C mark the area outside which the neural network is extrapolating. The diamond shapes in the left column mark the cases seen also on the right. The ITER simulations depicted here stem from both the training and the test set.}
    \label{fig:transfer1}
\end{figure}

\begin{figure}
    \centering
    \includegraphics[width=\textwidth]{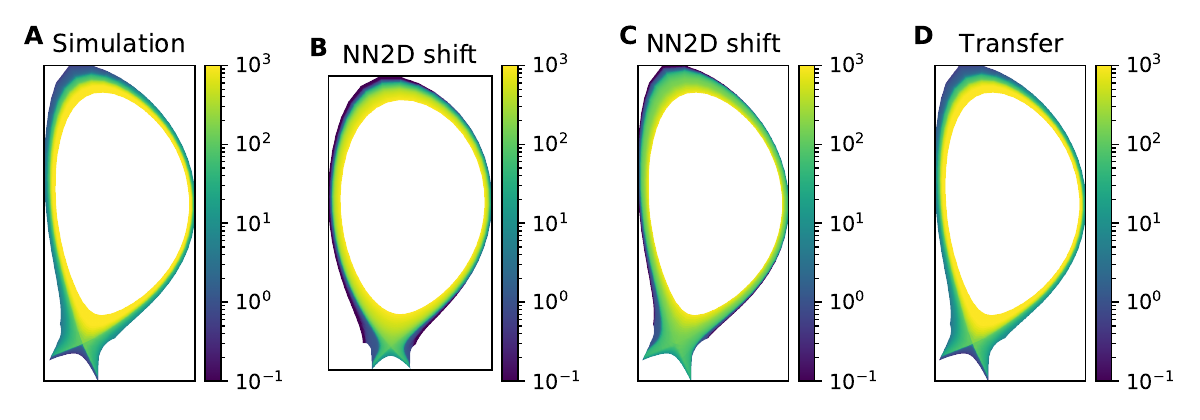}
    \caption{Electron temperature in one ITER simulation in the test set (A) and the corresponding predictions by the 
     NN2D fluid neutral model with rescaled gas puff (B), the same prediction but geometrically shaped to ITER geometry (C) and the transfer learned model (D).}
    \label{fig:transfer2D}
\end{figure}

\begin{figure}
    \centering
    \includegraphics[width=\textwidth]{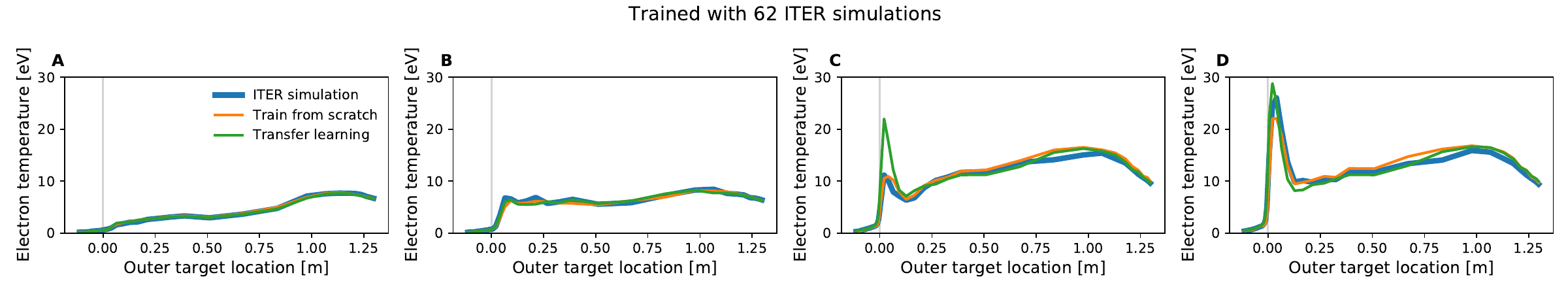}
    \includegraphics[width=\textwidth]{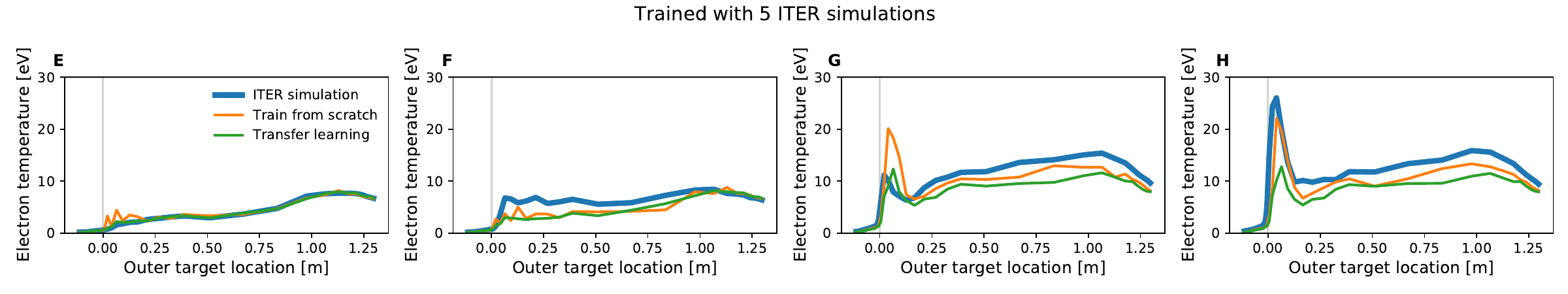}
    \caption{Comparison between the temperature profiles predicted by the transfer learned model (green) and the model trained from scratch (orange) at the outer target for four ITER simulations (blue) from the test set. The models are either trained with 62 (A-D) or 5 (E-H) ITER training simulations.}
    \label{fig:transfer1D}
\end{figure}

\begin{figure}
    \centering
    \includegraphics[width=\textwidth]{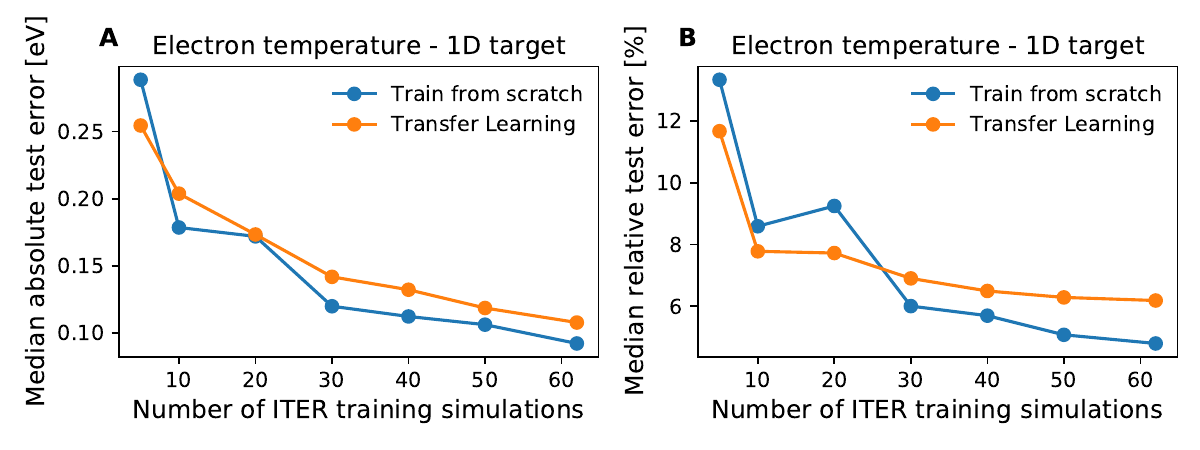}
    \caption{Median absolute (A) and relative (B) errors of the predicted temperatures at the outer target on the ITER test set for models trained with varying amount of ITER training simulations either from scratch (blue) or as transfer learning based on the previous fluid neutral surrogate (blue).}
    \label{fig:transfer_scaling}
\end{figure}

The surrogate models developed in the preceeding sections rely on a large dataset of reduced-fidelity simulations (Section~\ref{sec:dataset}).
The scope of the dataset and the reduced physics fidelity determine the applicability of the surrogate models.
While the previous section shows a rough first assessment of the validity of the final models for present day tokamaks, here a more in-depth comparison is made to determine the usefulness of the developed models for ITER.
\\
For this the surrogate model is compared against a set of SOLPS-ITER simulations from the ITER IMAS database (in the following referred to as ITER simulations).
These ITER simulations were already presented in \cite{pacher_impurity_2015,pitts_physics_2019,moulton_comparison_2021} and an exact list of the IMAS pulse numbers used here can be found in Table A.4 of \cite{dasbach_surrogate_2025}. 
The comparison is made purely for the electron temperature model but like in Section \ref{sec:multi} it is likely that the proposed error correction mechanisms work identically for all other variables.
These ITER simulations differ from our training data in several ways: ITER geometry (versus scaled JET), a kinetic neutral model, stricter convergence criteria, different gas puff/pump methods, helium ions, and neon instead of nitrogen seeding.
Only three parameters are independently varied in these simulations: input power, deuterium puff, and neon puff.
The helium inflow is fully correlated with power, so it does not count as an additional degree of freedom.
Assuming a 1:1 correspondence between neon and nitrogen puffing rates, the parameter space of the selected simulations overlaps with that of the surrogate model.
The previous NN2D model is used to create 10,000 predictions across a grid of deuterium and nitrogen puff values, fixing all other input parameters to match the ITER cases ($R=6.2\m$, $B=5.3\T$, $D_\perp=0.3\mmps$, $\chi_\perp=1.0\mmps$, $\Dcore=9.1\cdot10^{21}\atoms$, $\Pin=100\MW$).
Figure~\ref{fig:transfer1}(a) compares the predicted electron temperatures at the outer target separatrix against the ITER simulations.
For same gas puff values, the surrogate model predicts temperature significantly higher than in the ITER simulations, sometimes by several orders of magnitude. Figure~\ref{fig:transfer1}(b) compares the results for a fixed nitrogen/neon puff.
At low deuterium puff rates, the surrogate predicts a sheath-limited regime with target temperatures equal to the outer midplane temperature. At higher puff rates, the upstream temperature remains relatively constant (100–200 eV), whilst the target temperature decreases to \~10 eV and then further declines.
In the ITER simulations, target temperatures are consistently lower, but follow the same overall trend at almost an order of magnitude lower gas puff.
To align the surrogate to the ITER simulations, scaling factors for the input gas puffs can be introduced.
Due to the differences in the underlying physics the surrogate might need a higher gas throughput to elicit the same effect on the plasma.
Since even the highest-fidelity SOLPS-ITER simulations often have a drastically different throughput than experiments \cite{wiesen_plasma_2017}, this is an acceptable correction and similar approaches have been applied in previous code-code comparisons \cite{wiesen_control_2017}.
Applying the scaling factors shifts the surrogate predictions in gas puff space (Figures~\ref{fig:transfer1}c,d) which significantly improves the agreement with the ITER simulations.
However, this simple correction does not resolve all discrepancies: when additional observables (e.g., ion densities) are considered, no single consistent scaling can match all quantities simultaneously.
Even when considering only the electron temperature, a simple gas puff scaling can only correct the results at a few key locations at once, but not the entire spatial distribution as a whole.
Figure~\ref{fig:transfer2D}(a) shows the temperatures in one of the ITER simulations and Figure~\ref{fig:transfer2D}(b) the corresponding surrogate prediction after the gaspuff calibration shift. One obvious deviation is that the surrogate still predicts a scaled up JET case and not the actual ITER geometry.
Such differences can be corrected by using simple geometric transformations to map the solution from JET to ITER geometry Figure~\ref{fig:transfer2D}(c).
But even then large visible differences in the spatial temperature distributions persist.
A more powerful adaptation strategy is called transfer learning or fine-tuning \cite{geron_hands-machine_2019}.
In this method, the final layer in the surrogate model is replaced with a new layer of neurons, while all previous layers persists.
And while the weights of the connections in all the previous layers stay frozen, only the last layer is now trained on the ITER simulations.
The idea is that the model retains some of the learned physics from the previous database while having enough freedom to adapt to the new data.
To explore this method the ITER simulations are randomly split into a set of 62 training simulations for the transfer learning while 16 are withhold as test set.
Figures~\ref{fig:transfer1}(e,f) show the predictions of the updated model.
This model now offers the highest agreement with the ITER simulations, however some of the previous model behaviour is lost.
Most notably the temperatures predicted at the divertor target stagnate between 20-30eV for decreasing gas puffs.
This is not realistic as at some point a transition into a sheath-limited regime should occur similar as before.
Transfer learning as it is applied here is prone to this type of error, which is called \textit{catastrophic forgetting}.
Nevertheless, the model now achieves accurate temperature predictions across the whole spatial domain (Figure~\ref{fig:transfer2D}(d)).
Finally the question remains whether this transfer learning approach offers a benefit compared to training a new model from scratch solely on the ITER simulations.
In machine learning transfer learning is often employed to reduce the compute time it takes to train new models.
But for the simple models here training time is on the order of minutes, which is negligible compared to the time it takes to run even a single SOLPS-ITER simulation.
Therefore the key questions are whether the transfer learned model is more accurate or if it can reach the same levels of accuracy using fewer high-fidelity ITER simulations.
To test this, NN2D models were trained both via transfer learning and from scratch using varying numbers of ITER simulations. For models trained from scratch the hyperparameters were optimized via random search (see Table \ref{tab:NN2D-search}).
Using the full ITER training set, both approaches accurately predict outer target temperatures across different detachment levels (Figure \ref{fig:transfer1D}A–D). With only five ITER training simulations, both still capture trends in target profiles, though deviations increase visibly (Figure \ref{fig:transfer1D}E–H).
Median test errors decrease with more training simulations (Figure \ref{fig:transfer_scaling}). Models trained from scratch obtain slightly better scores at higher data counts, but differences are small, and transfer learning does not reduce the number of required high-fidelity simulations.
The test error values are notably lower than for the fluid neutral dataset in the previous sections, reflecting the smaller parameter space and limited temperature variation in ITER simulations. So even though the underlying physics of the simulations is more complex than in the previous sections one could argue that this dataset is more simple for the models to learn and reproduce.
This simplicity is further highlighted by the fact that models trained from scratch on ITER simulations achieve optimal performance with only two hidden layers (Table \ref{tab:NN2D-search}).
\\
In conclusion, the surrogate model trained on the fluid neutral simulations shows already very similar trends as present in kinetic neutral simulations. Transfer learning allows adapting the model to higher fidelity simulations but the tested approach does not yield a more accurate model than training solely on few high fidelity simulations.
But a definitive assessment is prohibited due to the small size of the high fidelity test set and more rigorous studies should evaluate the proposed methods further.

\newpage
\FloatBarrier
\section{Conclusions \& Discussion}\label{sec:discuss}
The presented study explored the development of a surrogate model for the two-dimensional scrape-off layer plasma. To this end, a database of SOLPS-ITER simulations with a reduced fidelity fluid neutral model was used. Based on this data, Gradient Boosting and neural network based models have been trained and evaluated.
While both machine learning methods were able to reproduce the results obtained in the simulations, the neural network models achieved higher accuracies.
\\
Training the neural network to output the electron temperature either solely at the divertor target or in the whole simulation domain, obtains similar levels of accuracy.
A similar comparison was made between a surrogate model constructed of independent networks for different plasma properties and a model that relies on a single network for multiple quantities.
Although the differences are small, the current recommendation is to train independent networks which each predict one plasma property in the whole simulation domain.
The developed surrogate models struggle with preserving the spatial gradients of plasma temperatures in low density simulations with high absolute temperature values. In such cases, deriving heat fluxes from the surrogate outputs leads to large errors.
The usage of designated models, which are trained to predict the heat fluxes themselves, avoids these errors.
\\
It was demonstrated that the developed surrogate model can reproduce an experimental scaling in ASDEX Upgrade and JET.
On the one hand this demonstrates the usefulness of cross-machine capable models.
On the other hand the models trained on the smaller ITER dataset are able to obtain accurate results with less than 60 training simulations. So for designated applications it might be worthwhile to decrease the scope of the developed surrogate models in favor of higher accuracy.
Transfer learning from the surrogate model trained on the fluid neutral dataset to the ITER dataset has shown to yield no improvements compared to developing a model on the ITER dataset from scratch. This might change when more complex higher fidelity datasets or more advanced network architectures are tested.
\\
The key to bring surrogate models of the tokamak scrape-off layer towards state-of-the-art accuracy and validity lies in decreasing the necessary number of training simulations and accelerating each individual simulation.
The work presented can act as a benchmark for more advanced model architectures and training strategies.
Possible pathways to explore further include active learning \cite{preuss_global_2018,bhosekar_advances_2018,geron_hands-machine_2019}, mixed fidelity models and the inclusion of experimental results.
At the moment it remains undecided whether one general cross-machine model or device specific models will come out ahead in the amount of simulations versus accuracy trade-off. As it stands both pathways seem like justifiable options.
To aid the process of creating a database of scrape-off layer simulations, more research should be conducted to find adaptive algorithms that change the numerical parameters of a simulation automatically.
This should also include more research into how the already developed surrogate can be used to provide initial plasma states for consecutive simulations.
\\
Overall, the present level of agreement with experimental scaling laws suggests that the surrogate models are already sufficiently reliable for fast scoping studies, enabling efficient exploration of operational spaces for next-generation devices such as ITER, DEMO or SPARC. Beyond this, their flexibility opens a wide range of applications, including rapid assessment of exhaust operational windows, coupling with core plasma models for integrated simulations, uncertainty propagation, and inverse optimization to match desired target profiles. In particular, the ability to evaluate large parameter spaces in near real time enables systematic studies to support the development of robust exhaust scenarios that ensure protection of plasma-facing components.

\FloatBarrier
\section{Data availability}\label{sec:data-availability}
The final trained models as used in the direct prediction method of Section~\ref{sec:loop} are available on Github: \href{https://github.com/sdasbach/solps-nn}{https://github.com/sdasbach/solps-nn}.
A compressed form of the training data described in Section \ref{sec:dataset} can be found on Zenodo \cite{dasbach_2026_19237127}.
The ITER models of Section \ref{sec:transfer} are work in progress and are available upon reasonable request or will be made available in revised form at a later stage.

\FloatBarrier
\section{Acknowledgements}\label{sec:acknowledge}
This work has been carried out within the framework of the EUROfusion Consortium, funded by the European Union via the Euratom Research and Training Programme (Grant Agreement No 101052200 — EUROfusion). Views and opinions expressed are however those of the author(s) only and do not necessarily reflect those of the European Union or the European Commission. Neither the European Union nor the European Commission can be held responsible for them.
\\\\
The views and opinions expressed herein do not necessarily reflect those of the ITER Organization. ITER is the nuclear facility INB 174. This work was conducted under the auspices of the ITER Scientist Fellow Network (ISFN).
\\\\
The authors gratefully acknowledges computing time on the supercomputer JURECA \cite{thornig_jureca_2021} at Forschungszentrum Jülich under grant no. solsur.
\\\\
This publication is part of the project \textit{Fusion Tokamak edge modeling} of the research programme \textit{Computing Time} that is (co-)funded by the Dutch Research Council (NWO). We acknowledge NWO for providing access to the Dutch National Supercomputer Snellius, hosted by SURF through the call for proposals 'Computing Time on National Computer Facilities'.

\newpage
\FloatBarrier
\appendix
\section{Pre-processing \& Hyperparameter optimization}\label{app:HPO}
For training a machine learning model it is often helpful if the data is first scaled to values close to zero.
For each of the eight simulation input parameters this is done by substracting the mean and dividing by the standard deviation $\hat{x}_{i}=\frac{x_i-\mu(\mathbf{x})}{\sigma(\mathbf{x})}$. For each parameter the scaled data then has mean 0 and standard deviation 1.
Because the parameters $\Dpuff$, $\Npuff$ and $\Dcore$ are varied on a logarithmic scale, the standardization is applied on the
logarithm (base 10) of these values. The standardized simulation parameters then constitute the input to the machine learning models in all previous sections.
\\
Regarding the pre-processing of the plasma states two factors have to be considered:
the drastically changing statistics at differing locations in the 2D domain (see e.g. Figure~\ref{fig:regimes}(b) and (c))
and the large variation between different simulations (seen in Figure~\ref{fig:regimes}(a)).
To address this a non-linear quantile transformation is used and applied independently for each grid cell in the simulation domain (different locations are treated as different features).
The quantile transformation (implemented in scikit-learn \cite{pedregosa_scikit-learn_2011})
estimates the cumulative distribution function of the data $F(\mathbf{x})$ and then applies this to each datapoint to transform them to a scale between 0 and 1. Then the quantile function $G^{-1}$ (inverse of the cumulative distribution function $G$) of a standard normal distribution is applied. Thus the scaled data points $\hat{x}_{i}=G^{-1}(F(x_i))$ follow almost exactly a normal distribution. This method is robust against outliers, but distorts correlations between datapoints.
In all sections the scaling procedure is fitted on the training simulations and applied to training and test simulations.

\begin{table}
    \centering
    \tiny
	\begin{tabular}{|c|c|c|c|c|}
		\hline
		  & NN2D & NNpos2D & NN1D & NN2D-ITER\tabularnewline
        \hline
        \hline
        Trials & 320 & 64 & 160 & 100\tabularnewline
		\hline
		Number of layers & [1,15] - $\mathbf{9}$ & [8,12] - $\mathbf{9}$ & [1,15] - $\mathbf{13}$ & [1,15] - $\mathbf{2}$\tabularnewline
		\hline
		Neurons per layer & [100,1500] - $\mathbf{1427}$ & [100,1500] - $\mathbf{456}$ & [100,2000]  - $\mathbf{1084}$ & [100,2000]  - $\mathbf{1313}$\tabularnewline
		\hline
		Learning rate & [$10^{-4}$,$5\cdot10^{-3}$] - $\mathbf{1.04\cdot10^{-4}}$ & $5\cdot10^{-4}$ & [$10^{-5}$,$5\cdot10^{-3}$] - $\mathbf{3.9\cdot10^{-4}}$ & [$10^{-4}$,$5\cdot10^{-3}$] - $\mathbf{1.4\cdot10^{-4}}$\tabularnewline
		\hline
		L2 regularization & [$10^{-5},10^{-2}$] - $\mathbf{3.4\cdot10^{-5}}$ & $5\cdot10^{-5}$ & [$10^{-5},10^{-2}$] - $\mathbf{1.5\cdot10^{-4}}$ & [$10^{-5},10^{-2}$] - $\mathbf{1.2\cdot10^{-3}}$\tabularnewline
		\hline
		Activation & \{Relu, Elu, Selu\} - \textbf{Selu} & Elu & \{Relu, Elu, Selu\} - \textbf{Elu} & \{Relu, Elu, Selu\} - \textbf{Elu}\tabularnewline
		\hline
		Mini-batch size & [20,200] - \textbf{37} & 500 & [20,200] - \textbf{178} & 32\tabularnewline
		\hline
		Batch normalization & \{True,False\} - \textbf{False} & False & \{True,False\} - \textbf{False} & \{True,False\} - \textbf{True}\tabularnewline
        \hline
        Loss function & MAE & MAE & MAE & MAE \tabularnewline
        \hline
        Optimizer & Adam & Adam & Adam  & Adam \tabularnewline
        \hline
        Patience & 50 & 5 & 50 & 50\tabularnewline
		\hline
	\end{tabular}
	\caption{Hyperparameters varied in the hyperparameter searches for the NN2D, NNpos2D and NN1D models in Section \ref{sec:2D} and the NN2D trained from scratch on all ITER simulation in Section \ref{sec:transfer}. $\left[...\right]$ denote ranges in which parameters are varied, while $\{...\}$ are discrete sets. The bold values are the optimal values found in the search. The learning rate and L2 regularization were sampled in logarithmic domain. Batch normalization was only tested with Relu and Elu activations. Each hidden layer has the same number of neurons. All other model hyperparameters remain at the default values set in TensorFlow. The NNpos2D-all-in-one model was not optimised but has 10 hidden layers with 300 neurons each, Mini-batch size of 4096, 10 epochs patience and is otherwise identical to the NNpos2D model.}\label{tab:NN2D-search}
\end{table}
\begin{table}
    \centering
    \small
	\begin{tabular}{|c|c|}
		\hline
		   & XGBoost2D\tabularnewline
        \hline
        \hline
        Trials & 50\tabularnewline
		\hline
		Tree depth & [5,20] - $\mathbf{10}$\tabularnewline
		\hline
		Learning rate & [$10^{-3},2\cdot10^{-1}$] - $\mathbf{0.11}$\tabularnewline
		\hline
		Number of Estimators & [200,1000] - $\mathbf{715}$\tabularnewline
        \hline
        Loss function & \{MAE,MSE\} - $\mathbf{MSE}$\tabularnewline
        \hline
        Subsample & [0.2,1.0] - $\mathbf{0.91}$\tabularnewline
		\hline
	\end{tabular}
	\caption{Hyperparameters varied in the hyperparameter search for the XGBoost2D model in Section \ref{sec:2D}. $\left[...\right]$ denote ranges in which parameters are varied, while $\{...\}$ are discrete sets. The bold values are the optimal values found in the search. The learning rate is sampled in logarithmic domain. All other model hyperparameters remain at the default values set in XGBoost.}\label{tab:XGBoost-search}
\end{table}

\newpage
\FloatBarrier
\section{Tokamak parameters}\label{sec:goldston}
\begin{table}
	\centering
    \small
	\begin{tabular}{|c|c|c|c|c|}
		\hline 
		& AUG & JET & ITER\tabularnewline
		\hline 
		\hline 
		$R$ & $1.6\m$ & $2.9\m$ & $6.2\m$\tabularnewline
		\hline 
		$B$ & $2.5\T$ & $2.5\T$ & $5.3\T$\tabularnewline
		\hline 
		$\Pin$ & $10.7\MW$ & $14\MW$ & $100\MW$\tabularnewline
		\hline 
		$\chi_\perp$ & $0.6\ \mathrm{m}^2/\mathrm{s}$ & $0.5\ \mathrm{m}^2/\mathrm{s}$ & $1.0\ \mathrm{m}^2/\mathrm{s}$\tabularnewline
		\hline 
		$D_\perp$ & $0.5\ \mathrm{m}^2/\mathrm{s}$ & $0.1\ \mathrm{m}^2/\mathrm{s}$ & $0.3\ \mathrm{m}^2/\mathrm{s}$\tabularnewline
		\hline 
		$\Dcore$ & $1 \cdot 10^{21}\ \mathrm{atoms/s}$ & $1.5 \cdot 10^{21}\ \mathrm{atoms/s}$ & $9.1 \cdot 10^{21}\ \mathrm{atoms/s}$\tabularnewline
		\hline 
		$n_\mathrm{GW}$ & $1.44\cdot10^{20}\mathrm{m}^{-3}$ & $9.82\cdot10^{19}\mathrm{m}^{-3}$ & $1.19\cdot10^{20}\mathrm{m}^{-3}$\tabularnewline
		\hline 
	\end{tabular}
	\caption{Surrogate model inputs used to represent AUG and JET in Section \ref{sec:physics} and ITER in \ref{sec:transfer}. The transport coefficients were selected as in \cite{rozhansky_multi-machine_2021}, $\Dcore$ as in \cite{xiang_modeling_2017} (AUG) and \cite{kaveeva_solps-iter_2021} (JET). All other parameters are taken from Table 2 in \cite{goldston_new_2017}.}
	\label{tab:goldston}
\end{table}

\begin{figure}
    \centering
    \includegraphics[width=0.5\textwidth]{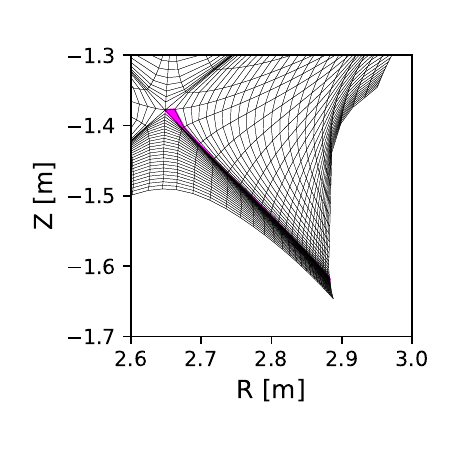}
    \caption{Highlighted in purple are the grid cells used for the volume integral to calculate the divertor impurity concentration $c_\mathrm{N,div} = \int(n_\mathrm{N} + n_\mathrm{N^+} + \dots + n_\mathrm{N^{7+}}) \mathrm{dV} / \int n_\mathrm{e}\mathrm{dV}$. The volume of each grid cell is given by its area in the poloidal cross-section multiplied with $2\pi$ and the major radius of each grid cell.}
    \label{fig:impurity-zone}
\end{figure}

\FloatBarrier

\printbibliography

\end{document}